# Super-resolution microscopy via ptychographic structured modulation of a diffuser

PENGMING SONG,[1,3] SHAOWEI JIANG,[2,3] HE ZHANG,[2,3] ZICHAO BIAN,[2] CHENGFEI GUO,[2] KAZUNORI HOSHINO,[2] AND GUOAN ZHENG,[1,2,*]

[1]Department of Electrical and Computer Engineering, University of Connecticut, Storrs, CT, 06269, USA
[2]Department of Biomedical Engineering, University of Connecticut, Storrs, CT, 06269, USA
[3]These authors contributed equally to this work
*Corresponding author: guoan.zheng@uconn.edu

**We report a new coherent imaging technique, termed ptychographic structured modulation (PSM), for quantitative super-resolution microscopy. In this technique, we place a thin diffuser (i.e., a scattering lens) in between the sample and the objective lens to modulate the complex light waves from the object. The otherwise inaccessible high-resolution object information can thus be encoded into the captured images. We then employ a ptychographic phase retrieval process to jointly recover the exit wavefront of the complex object and the unknown diffuser profile. Unlike the illumination-based super-resolution approach, the recovered image of our approach depends upon how the complex wavefront exits the sample – not enters it. Therefore, the sample thickness becomes irrelevant during reconstruction. After recovery, we can propagate the super-resolution complex wavefront to any position along the optical axis. We validate our approach using a resolution target, a quantitative phase target, a two-layer sample, and a thick PDMS sample. We demonstrate a 4.5-fold resolution gain over the diffraction limit. We also show that a 4-fold resolution gain can be achieved with as few as ~30 images. The reported approach may provide a quantitative super-resolution strategy for coherent light, X-ray, and electron imaging.**

*OCIS codes: (180.0180) Microscopy; (100.5070) Phase retrieval; (150.6910) Three-dimensional sensing*

http://dx.doi.org/

In structured illumination microscopy, non-uniform illumination patterns are used to modulate the otherwise inaccessible object information into the passband of the optical system [1-4]. Instead of using illumination patterns, modulation of the object information can also be performed at the detection path using disordered media [5-12]. In the past years, it has been shown that the disordered media can serve as a scattering lens for coherent light wave modulation. Resolution beyond the diffraction limit can be achieved via wavefront shaping or transmission matrix characterization [5-12].

Inspired by the concept of the scattering lens, we report a new coherent super-resolution imaging technique, termed Ptychographic Structured Modulation (PSM), in this letter. In the PSM technique, we place a thin diffuser (i.e., a scattering lens) in between the sample and the objective lens to modulate the complex light waves. We then scan the diffuser to different lateral positions and acquire the modulated images through the objective lens. The acquired images are used to recover the super-resolution exit wavefront of the complex object and the diffuser profile using a ptychographic phase retrieval process [13-16].

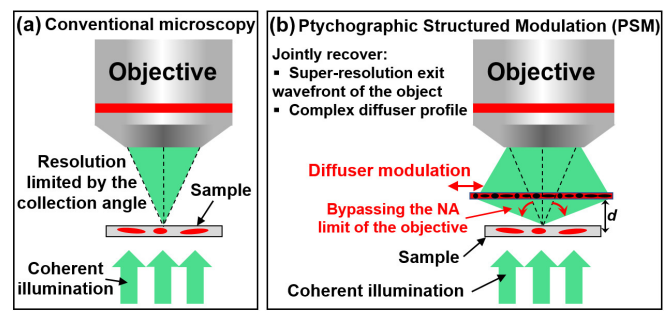

Fig. 1. Comparison between the regular coherent microscope platform (a) and the proposed PSM concept (b). In the PSM approach, we use a thin diffuser to modulate the light waves from the object. The diffuser is then scanned to different lateral positions and the captured images are used to recover the super-resolution exit wavefront of the object and the diffuser profile. With the recovered exit wavefront, we can digitally propagate it to different z positions to get the object images. The final achievable resolution is limited by the smallest feature size of the diffuser, which modulates the otherwise inaccessible object information into the passband of the objective.

Figure 1 shows the comparison between a regular coherent microscopy imaging setup and the reported PSM concept. As shown in Fig. 1(b), the high-resolution object information otherwise inaccessible is now encoded in the captured images in the PSM approach. The final achievable resolution is not limited by the numerical aperture (NA) of the objective lens. Instead, it is limited by the feature size of the thin diffuser. In our experiment, we demonstrate a 4.5-fold resolution gain beyond the diffraction limit of the employed objective lens. We also show that, a 4-fold resolution gain can be achieved with as few as ~30 images.

Drawing connections and distinctions between the proposed approach and its related imaging modalities helps to clarify its operation. In the turbid lens imaging technique developed by Choi et al. [7, 8, 12], the transmission matrix of the diffuser is measured and then used to recover the super-resolution complex object. In our approach, we use a thin diffuser for light wave modulation. The transmission matrix of our diffuser can be approximated via a diagonal matrix. As such, we can model the interaction between the light waves and the diffuser using point-wise multiplication. Instead of directly measuring the transmission matrix in PSM, we employ the ptychographic phase retrieval process to recover the super-resolution exit wavefront of the complex object and the complex diffuser profile.

The reported PSM approach also shares its roots with super-resolution ptychography [13], near-field ptychography [14-18], and Fourier ptychography (FP) [19]. In super-resolution ptychography, speckle patterns are used to improve the achievable resolution. The illumination probe, however, is confined to a limited region in the object space, leading to a large number of image acquisitions. Our approach, on the other hand, employs a scattering lens to modulate the light waves for the entire field of view (instead of a confined region). In our demonstration, as few as ~30 images can be used to recover the super-resolution object with 4-fold resolution gain beyond the diffraction limit.

Near-field ptychography uses a translated speckle pattern to modulate the object over the entire field of view. The key difference between our approach and near-field ptychography is the use of a scattering lens for super-resolution imaging. By placing the diffuser layer in between the object and the detection optics, the otherwise-lost high-resolution object information can be now encoded into the captured images [7, 8, 12]. We can, therefore, substantially improve the resolution beyond the diffraction limit of the employed objective lens.

FP illuminates the object with angle-varied plane waves and recovers the super-resolution complex object profile. Thin object assumption is needed in FP as well as in super-resolution ptychography. Only under this assumption, the interaction between the illumination wave and the object can be approximated via point-wise multiplication. Unlike the illumination-based implementations in FP, the reported PSM approach modulates the object light waves at the detection path. The thin object requirement in FP and ptychography is converted into thin diffuser requirement in PSM. The recovered exit wavefront of PSM depends upon how the light field exits the sample – not enters it. Therefore, the sample thickness becomes irrelevant during reconstruction. After recovery, we can digitally propagate the super-resolution complex wavefront to any plane for 3D holographic refocusing.

The reported PSM approach can also be used in coherent X-ray and electron microscope. A thin diffusing layer can be added in between the object and the objective lens to modulate the X-ray photons and electrons otherwise inaccessible by the systems. It can improve the imaging resolution and recover quantitative object phase in current X-ray and electron microscopes without major hardware modifications.

The forward imaging model of the PSM approach in Fig. 1(b) can be expressed as

$$I_j(x,y) = \left| \left[ \left( W(x,y) * PSF_{free}(d) \right) \cdot D(x - x_j, y - y_j) \right] * iFT\{CTF(k_x, k_y)\} \right|^2, \quad (1)$$

where $I_j(x,y)$ is the $j^{th}$ intensity measurement $(j = 1,2,...,J)$, $W(x,y)$ is the complex exit wavefront of the object, $D(x,y)$ is the complex profile of the diffuser, $(x_j, y_j)$ is the $j^{th}$ positional shift of the diffuser, $iFT$ stands for inverse Fourier transform, '·' stands for point-wise multiplication, and '*' denotes convolution operation. In Eq. (1), '$d$' is the distance between the exit wavefront and the diffuser. We use $PSF_{free}(d)$ to model the point spread function (PSF) for free-space propagation over distance $d$. Similarly, $CTF(k_x, k_y)$ is the defocus coherent transfer function (CTF) with a defocus distance of '-$d$'.

With the recovered complex exit wavefront $W(x,y)$, we can then digitally propagate it to different axial positions via:
$$O_z(x,y) = W(x,y) * PSF_{free}(z), \quad (2)$$
where $O_z(x,y)$ is the recovered object profile at the '$z$' position.

---

**Recovery process of the Ptychographic Structured Modulation (PSM) approach**

**Input**: Raw image sequence $I_j(j = 1,2,\cdots,J)$
**Output**: Super-resolution exit wavefront $W(x,y)$ and the diffuser profile $D(x,y)$

1  Initialize $W(x,y)$, $D(x,y)$, and the defocus $CTF$
2  **for** $n = 1: N$  (different iteration loops)
3   **for** $j = 1: J$ (different captured images by scanning the diffuser)
4    $W'(x,y) = W(x,y) * PSF_{free}(d)$  % Propagate the wavefront to the diffuser
5    $D_j(x,y) = D(x - x_j, y - y_j)$  % Shift the diffuser to different x-y positions
6    $\phi_j(x,y) = W'(x,y) \cdot D_j(x,y)$
7    $\Phi_j(k_x, k_y) = \text{FFT}\left(\phi_j(x,y)\right)$
8    $\Psi_j(k_x, k_y) = \Phi_j(k_x, k_y) \cdot CTF$ % Spectrum filtering via the defocus CTF
9    $\psi_j(x,y) = \text{IFFT}\left(\Psi_j(k_x, k_y)\right)$
10   $\Psi'_j(k_x, k_y) = \text{FFT}\left(\psi_j(x,y)/|\psi_j(x,y)| \cdot \sqrt{I_j}\right)$ % Amplitude replacement
11   Update the spectrum $\Phi_j(k_x, k_y)$: $\Phi'_j = \Phi_j + \beta_\Phi \frac{conj(CTF)(\Psi'_j - \Psi_j)}{|CTF|^2_{max}}$
12   $\phi'_j(x,y) = \text{IFFT}\left(\Phi'_j(k_x, k_y)\right)$
13   Update $W'(x,y)$: $W' = W' + \frac{conj(D_j)(\phi'_j - \phi_j)}{(1-\alpha_{obj})|D_j|^2 + \alpha_{obj}|D_j|^2_{max}}$
14   Update $D_j(x,y)$: $D_j = D_j + \frac{conj(O')(\phi'_j - \phi_j)}{(1-\alpha_D)|O'|^2 + \alpha_D|O'|^2_{max}}$
15   $D(x,y) = D_j(x + x_j, y + y_j)$  % Shift back the updated diffuser profile
16   $W(x,y) = W'(x,y) * PSF_{free}(-d)$ % Propagate back to the object plane
17   Add Nesterov momentum
18  **end**
19 **end**

Fig. 2. The recovery process of the PSM approach, where we acquire images by scanning the diffuser to different positions. The intensity measurements are then used to recover the exit wavefront and the diffuser profile.

Based on all captured images $I_j$ with the diffuser scanned to different lateral positions $(x_j, y_j)$s, we aim to recover the complex exit wavefront of the object $W(x,y)$ and the diffuser profile $D(x,y)$. The recovery process is shown in Fig. 2. We first initialize the amplitude of the exit wavefront by averaging all measurements. The diffuser profile is initialized to an all-one matrix. The defocus CTF is initialized based on an estimate of the distance $d$ between the exit wavefront and the diffuser. In the reconstruction process, we first propagate the object to the diffuser plane and obtain $W'$ in line 4. We then multiply $W'$ with the shifted diffuser to obtain the exit wave $\phi_j$ in line 6. The exit wave is low pass filtered in the Fourier domain to get $\psi_j(x,y)$ in line 9. The amplitude of $\psi_j(x,y)$ is then replaced by the $j^{th}$ measurement $\sqrt{I_j}$ while the phase is kept unchanged. The Fourier spectrum of the exit wave $\phi_j$ is updated in the Fourier domain in line 11 [20]. Based on the updated exit wave $\phi'_j$, we then update the wavefront and the diffuser profiles in lines 13-14 [21]. We also add Nesterov momentum to accelerate the convergence speed in our implementation [21]. The processing time for 100 raw images with 1024 by 1024 pixels each is ~1 minute for 25 iterations using a Dell XPS 8930 desktop computer.

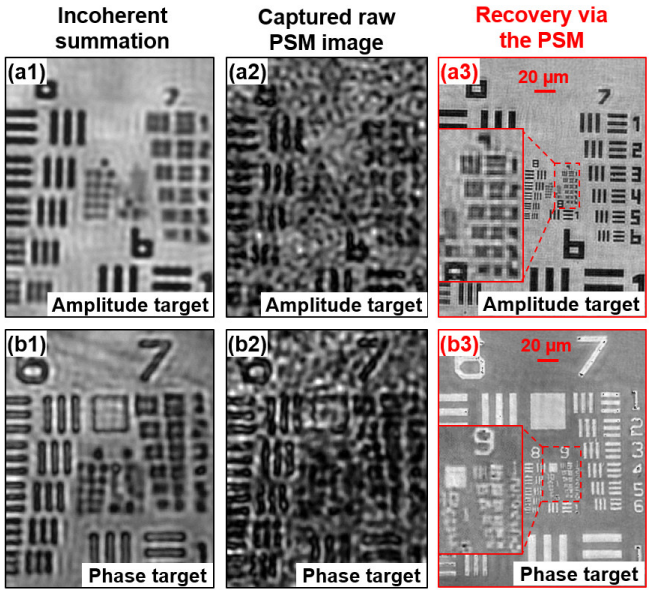

Fig. 3. Validation of the super-resolution capability of the PSM approach using an amplitude (a) and a phase (b) target. The incoherent summation of all captured images of the amplitude (a1) and the phase (b1) target. The raw PSM images of the amplitude (a2) and the phase (b2) target. The recovered super-resolution images of the amplitude (a3) and the phase (b3) target based on 864 raw images. In the recovered images, we can resolve the half-pitch line width of group 9, element 1, achieving ~4.5-fold resolution gain over the diffraction limit of the employed objective lens.

We first validate the super-resolution imaging capability of the PSM approach in Fig. 3. In this experiment, we use a microscope platform with a 2X, 0.055 NA objective lens (Mitutoyo Plan Apo) for image acquisition. The light source is a 532-nm fiber coupled laser. The sample is an amplitude resolution target in Fig. 3(a) and a phase target in Fig. 3(b). The diffuser is made by coating ~1-μm microspheres on a cover slip. The distance between the diffuser and the sample is about 0.5 mm. We use two mechanical stages (Applied Scientific Instrumentation LS-50) to scan the diffuser to different x-y positions and acquire the corresponding images. The positional shift of the diffuser is about 2-4 pixels in between adjacent acquisitions. Figures 3(a1) and 3(b1) show the incoherent summations of all acquired images. The diffraction-limited resolution is 4.38 μm half-pitch linewidth, corresponding to group 6, element 6. Figures 3(a2) and 3(b2) show the captured raw images of the resolution targets, where the speckle feature comes from the diffuser modulation. Figures 3(a3) and 3(b3) show our recovery, where we can resolve 0.98-μm linewidth from group 9, element 1 of the resolution targets. The resolution gain is 4.5-fold over the diffraction limit. The final NA of the PSM approach is determined by the spatial frequency content of the diffuser profile added with the NA of the employed objective lens. The current achievable resolution is limited by the feature size (1-1.5 μm) of the diffuser (the spatial frequency content of the diffuser profile corresponds to an NA of ~0.22). One can, for example, use $TiO_2$ nanoparticles to make a diffuser with substantially stronger modulation capability [5-12].

In the second experiment, we validate the quantitative imaging nature of the PSM approach. A quantitative phase target (Benchmark QPT) is used as the object. Figure 4(a) shows the captured raw image through diffuser modulation. Figure 4(b) shows the recovered phase using the PSM approach. The line profile across the red dash arc in Fig. 4(b) is plotted in Fig. 4(c). The recovered phase is in a good agreement with the ground-truth height of the phase target.

In the third experiment, we investigate the number of raw images needed for our recovery. Once we recover the complex diffuser profile, we can substantially reduce the number of images for reconstruction. Figure 5 shows the recovered results using different numbers of acquired images. We can see that a 4-fold resolution gain can be achieved with as few as ~30 images (resolving the linewidth of group 8, element 6). As shown in Fig. 5, it is the image quality (signal to noise ratio) becoming better with the increased number of images, resulting in increased visibility of smaller features.

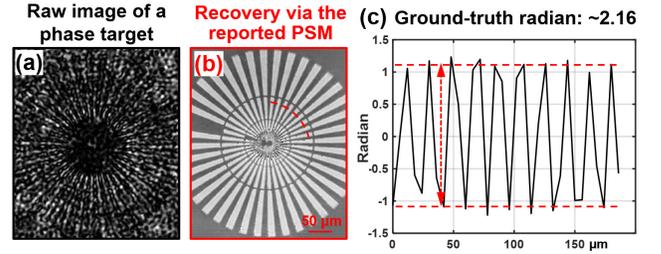

Fig. 4. Validating the quantitative imaging nature of the PSM approach. (a) The captured raw image through the diffuser. (b) The recovered phase image based on 864 raw images. (c) The line trace of the red arc in (b).

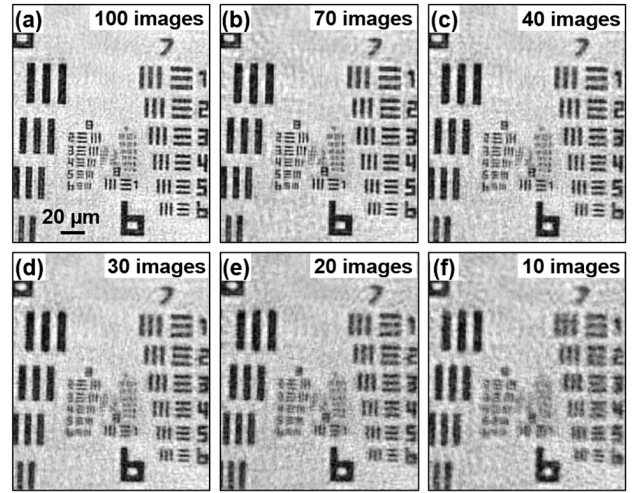

Fig. 5. Reconstruction with different numbers of raw images. (a) 100 images. (b) 70 images. (c) 40 images. (d) 30 images. (e) 20 images. (f) 10 images. We can achieve 4-fold resolution gain with as few as 30 images.

In the fourth experiment, we test the PSM approach with a two-layer biological sample and a thick polydimethylsiloxane (PDMS) sample mixed with microspheres. In Fig. 6, we place two pathology sections (Oesophagus cancer slides) together and they are separated by two coverslips. Figure 6(a1) and 6(a2) show the recovered amplitude and phase of the complex wavefront exiting the two-layer sample. Figure 6(b1) and 6(b2) shows the recovered object amplitude after digitally propagating to z = 180 μm and z = 500 μm. Top layer is in focus in Fig. 6(b1) and bottom layer is in focus in Fig. 6(b2). Visualization 1 shows the digital propagating process of the recovered wavefront. Similarly, Fig. 7 shows results of a thick PDMS sample mixed with microspheres. Figure 7(a) shows the digitally propagated wavefront at different axial planes (Visualization 2). Figure 7(b) shows the recovered 3D positions of microspheres by locating the intensity minimums of the back-propagated wavefront at different z positions.

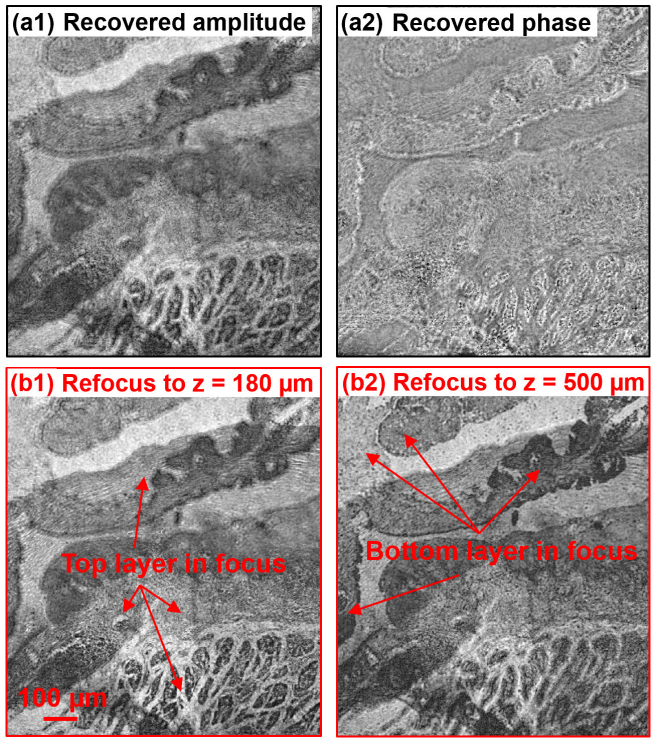

Fig. 6. (Visualization 1) Test the PSM approach using a two-layer object. (a) The recovered amplitude and phase the two-layer object. (b) Digital propagation of the recovered complex wavefront to two different layers.

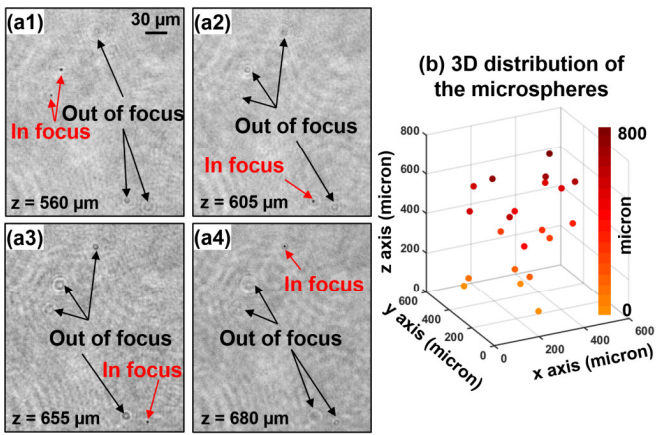

Fig. 7. (Visualization 2) Test the PSM approach using a thick PDMS sample mixed with microspheres. (a) Back-propagated object amplitude at 4 axial positions. (b) The recovered 3D positions of microspheres by locating the intensity minimums of the back-propagated wavefront at different z positions.

In summary, we report a coherent imaging technique, termed ptychographic structured modulation, for quantitative super-resolution microscopy. The reported PSM technique has several advantages. First, it can bypass the diffraction limit of the employed objective lens. We demonstrate a 4.5-fold resolution gain over the diffraction limit. We also show that a 4-fold resolution gain can be achieved with as few as ~30 images. Second, different from the structured illumination technique, the reported PSM modulates the wavefront at the detection path and it recovers the complex wavefront exiting the sample. Thin sample assumption plagued in regular ptychography and FP is no longer an issue in PSM. Third, the reported platform provides the true quantitative contrast of the complex object. It may provide a quantitative super-resolution strategy for coherent light, X-ray, and electron imaging. Finally, we also note that the PSM technique can also be implemented in a lensless microscopy platform [18] and the result will be presented elsewhere.

**Funding.** This work is supported by National Science Foundation 1510077 and National Institute of Health R03EB022144.